\documentclass[referee]{chjaa}

\usepackage{graphicx,times}             
\input{epsf.sty}                        
\input{psfig.sty}                       
\usepackage{latexsym}

\begin{document}

\title{High spatial resolution X-ray spectroscopy of SNR Cassiopeia A with {\sl Chandra}}

   \volnopage{Vol.0 (200x) No.0, 000--000}      
   \setcounter{page}{1}           

   \author{Xue-Juan Yang
      \inst{1,2}\mailto{yangxj@mail.ihep.ac.cn}
   \and Fang-Jun Lu
      \inst{1}
   \and Li Chen
      \inst{2}
      }

   \institute{Particle Astrophysics Center, Institute of High Energy Physics,
Chinese Academy of Sciences, Beijing 100049, China\\
             \email{yangxj@mail.ihep.ac.cn}
        \and
             Department of Astronomy, Beijing Normal University, Beijing
100875, China\\
          }

   \date{Received~~2007 Oct. 10; accepted~~year~~month day}

   \abstract{
We present high spatial resolution X-ray spectroscopy of supernova
remnant Cassiopeia A with the {\sl Chandra} observations. The X-ray
emitting region of this remnant was divided into 38 $\times$ 34
pixels with a scale of 10$\arcsec$ $\times$ 10$\arcsec$ each.
Spectra of 960 pixels were created and fitted with an absorbed two
component non-equilibrium ionization model. With the spectral
analysis results we obtained maps of absorbing column density,
temperatures, ionization ages, and the abundances for Ne, Mg, Si, S,
Ca and Fe. The Si, S and possibly Ca abundance maps show obviously
jet structures, while Fe doesn't follow the jet but seems to be
distributed perpendicular to it. In the range of about two orders of
magnitude, the abundances of Si, S and Ca show tight correlations
between each other, suggesting them to be ejecta from explosive
O-burning and incomplete Si-burning. Meanwhile, Ne abundance is well
correlated with that of Mg, indicating them to be the ashes of
explosive C/Ne burning. The Fe abundance is positively correlated
with that of Si when Si abundance is lower than 3 solar abundances,
but a negative correlation appears when the Si abundance is higher.
We suggest that such a two phase correlation is the results of
different ways in which Fe is synthesized.
   \keywords{ISM: supernova remnants -- ISM: individual: Cassiopeia A}
   }

   \authorrunning{X.-J. Yang et al.}            
   \titlerunning{{\sl Chandra} observation of Cas A}  
   \maketitle

%

\section{Introduction}

The young core-collapse supernova remnant (SNR) Cassiopeia A (Cas A)
is a perfect laboratory for studying the ejecta and shock in
remnants, for its age and distance are well-determined and it is
very bright across the whole electromagnetic spectrum. The age of
Cas A is around 300 years (Thorstensen et al. 2001) and the distance
about 3.4 kpc (Reed et al. 1995). It appears today as a nearly
circular 3 arcmin diameter bright ring, with a low surface
brightness 5 arcmin diameter radio and X-ray plateau. As a young
SNR, it is believed that thermal X-ray emission mainly comes from
the forward shocked interstellar medium (ISM) (Shklovsky 1973) and
the reverse shocked supernova (SN) ejecta (McKee 1974). The X-ray
image of Cas A is consistent with this scenario as suggested by
Gotthelf et al. (2001). They discovered a thin, bright X-ray wisp
that is interpreted as the forward shock, and a sharp rise in radio
and X-ray line emissivity at the inner edge of the bright ring,
which is associated with the reverse shock.

Numerical models (e.g., Woosley \& Weaver 1995; Thielemann et al.
1996) predict that nucleosynthesis in core-collapse supernovae (SNe)
occurs in an ``onionskin'' manner. Explosive Si burning occurs near
the core, where the shock temperatures is the highest. This process
would burn up Si and form ejecta dominated by $^{56}$Ni, which
decays to $^{56}$Co and finally $^{56}$Fe. Further out, as the shock
temperature decreases, Si burning is incomplete and the main
products include not only Fe, but also much Si, S, Ar and Ca. Even
further out, explosive O burning occurs, leading to a composition
dominated by O and Si with very little or no Fe. At the outmost the
explosive Ne/C burning occurs and forms mostly O. Willingale et al.
(2002) did a spectral mapping of Cas A using the data collected by
the XMM-Newton X-ray observatory ({\sl XMM-Newton}). They found that
the distributions of Si, S, Ar and Ca are very similar to each other
but distinct from Ne and Mg. This supports the above explosive
nucleosynthesis network. However, Hughes et al. (2000) and
Willingale et al. (2002) showed that the Fe-rich ejecta lie outside
the Si-rich material, using data from the Chandra X-ray observatory
({\sl Chandra}) and {\sl XMM-Newton} respectively. It was concluded
that the ejecta in Cas A have undergone a spatial inversion of the
explosive O- and Si-burning products. That is to say, the materials
from the predicted ``onion'' layers have mixed together. Hughes et
al. (2000) proposed this inversion to be the result of
neutrino-driven convection during the initiation of the SN
explosion.

{\sl Chandra} performed a 1 Ms observation on Cas A as a Very Large
Project (VLP), which was largely motivated by the two papers: Laming
\& Hwang (2003) and Hwang \& Laming (2003), focusing on the
suggested jet structure of the ejecta and the iron synthesis. Hwang
et al. (2004) showed the first sight of the observation. A jet
structure of the Si-rich ejecta was clearly found in the enhanced
Si-K emission line image. So it is proposed that Cas A was formed by
an asymmetric explosion, which is also indicated by the observed
kinematics and the high $^{44}$Ti yield (Vink 2004). Fe-rich ejecta,
however, was not found in the jet area, but perpendicular to it
instead. This contradicts to jet-induced explosion (Khokhlov et al.
1999), but might be the case for collapsars (Nagataki et al. 2003;
Zhang et al. 2004). The explosion energy for Cas A was about $2 \sim
4 \times 10^{51}$ ergs, suggesting it to be a normal SN (Laming \&
Hwang 2003). Cas A thus provides strong evidence that jets may also
be produced by normal SNe (Hwang et al. 2004).

In this paper, we present high resolution X-ray spectroscopy of Cas
A using the {\sl Chandra} observations. Willingale et al. (2002) did
a similar work with the {\sl XMM-Newton} data. Here we would like to
cross-check their results. In our work, we use a smaller pixel size
(10$\arcsec$ $\times$10$\arcsec$, 1/4 of that used by Willingale et
al. (2002)), which is closer to the typical size of the ejecta
knots of Cas A. Another point is that the so-called  ``jet'' region
is included in the analysis, while not in Willingale et al. (2002).
The high spatial and spectral resolutions of {\sl Chandra} together with
the abundant archived data permit us to study this famous SNR in
detail. In $\S$ 2, we briefly describe the observation and data
reduction. In $\S$ 3, we show the results, while some discussion in
$\S$ 4, and summary in $\S$ 5.

\section{Observation and Data Reduction}

Cas A was observed by {\sl Chandra} for about 1 Ms in 2004 as a VLP.
The details of the observations were described by Hwang et al.
(2004). In this paper, we used data from two deep ACIS exposure
segments of the VLP. One was performed from April 14 to 16
(Observation ID: 4638) and the other from May 25 to 26 (Observation
ID: 4639), with the exposure time of about 167 ks and 80 ks
respectively. The reasons that we used two segments are to ensure
good statistic for the spatially resolved analysis and to reduce the
mis-calibration effects, by using data collected in different
epochs.

The X-ray data were analyzed using CIAO software package (version
3.3, with the CALDB version 3.2.1 and the ATOMDB version 1.3.1). In
order to do spatially resolved analysis, the X-ray emission region
was divided into pixels with size about 10$\arcsec$ $\times$
10$\arcsec$, corresponding to a grid of 38 $\times$ 34. Figure 1
shows the image for observation ID 4639 with the grid overlaid. We
retained events within energies 0.5-10 keV using the archived level
2 event file and created spectra for 960 pixels that contain at
least 3000 counts. The frequency distribution of the counts in each
pixel is shown in Figure 2 (left panel). We can see that most pixels
contain more than 10000 counts, which allows statistical reliable
spectral analysis. The background spectrum was created from the
off-source region.

The spectral analysis was performed using the XSPEC (version 11.2)
package (Arnaud 1996). The spectra for each pixel of the two
observations were jointly fitted with two non-equilibrium ionization
components (VNEI, Borkowski et al. 2001). The free parameters are
the temperatures, emission measures, ionization ages ($\tau$={\sl
n$_e$t}) and abundances of O, Ne, Mg, Si, S, Ca and Fe for each of
the two components. The abundances are in units of solar abundances
given by Anders \& Grevesse (1989). We also introduced a uniform
redshift for each component to study the dynamics. The WABS model
(Morrison \& MacCammon 1983) was included to take care of
interstellar photo-electric absorption. A Gaussian line was added at
the energy of 3.4 keV to account for the Ar line emission, which is
not included in the VNEI model in XSPEC version 11.2 we used.

Here we note two points. One is that the nonthermal emission
contributes to the 4 $\sim$ 6 keV continuum without any doubt, which
could be as high as 25$\%$ in Cas A (Willingale et al. 2002).
However, since its X-ray emission is dominated by the thermal
component (Laming 2001; Bleeker et al. 2001) and we are mainly
interested in the study of line emission in this ppaer, we believe
that our model is appropriate. The same spectral model was also used
by Willingale et al. (2002). The other is that Cas A is an O-rich
SNR, so that O contributes a significant fraction of the electrons
(Vink et al. 1996; Willingale et al. 2002). In this case, the O
abundance is coupled with the emission measure. Here we set the O
abundances of both components free in the fitting process, but we
will not discuss about the O abundance in this paper as it might be
contaminated.

\section{Results}

The frequency distribution of the 960 reduced ${\chi}^2$ values of
the spectral fits is given in Figure 2 (right panel). It has a peak
around 0.7, which suggests our fitting results to be acceptable
statistically. In Figure 3, we give the spectra along with the
fitting residuals of several typical regions (marked in Figure 1).
The emission lines are also marked. Region A is located at the
outmost of the jet in the northeast. It has very strong Si, S, Ar
and Ca lines, but the Fe-K line is absent. Therefore it should be
dominated by O-burning products, but we cannot rule out the
incomplete Si-burning, as there seem to be the Fe-L lines around 1
keV (c.f. $\S$ 1). Region B is from the counter-jet, which is very
strong in Si and S. The Fe-K line is also clearly shown. This is
very consistent with the incomplete Si-burning yield out. Region C
is the Fe-rich region from the southeast. From the spectra we can
see very strong Fe-K and Fe-L lines, which are believed to be from
complete explosive Si-burning. Region D is located at the out rim,
which is weak in all the emission lines of the main elements in Cas
A and dominated by the continuum (c.f. Hughes et al. 2000, Figure
3). These spectra show that the X-ray properties change
significantly across the remnant. In the following we present the
statistical studies of the spectral fitting results.

(I) We got the spatial distribution of the absorbing column density
(c.f. Fig. 4), as well as temperatures and ionization ages for the
cool and hot components (c.f. Fig. 5). In order to show the large
scale structures more clearly, we smoothed all these maps with a
two-dimension Gaussian with FWHM 40$\arcsec$ $\times$ 40$\arcsec$,
and overplotted the smoothed contours on these maps. The column
density of the west part is higher than that of the east. This is
consistent with the former studies (Keohane et al. 1996; Willingale
et al. 2002), in which it was suggested that the absorption to the
west is higher due to the interaction of the remnant with the
molecular cloud. The column density lies in the range
$0.8-1.5\times10^{22}$ cm$^{-2}$ with the mean value of
1.19$\times10^{22}$ cm$^{-2}$. The typical statistical error is
$\sim 5\%$ at 90\% confidence level. The column density is a little
smaller than 1.50$\times10^{22}$ cm$^{-2}$ derived by Willingale et
al. (2002), but consistent with that from Keohane et al. (1996)
given by the equivalent width of HI and OH ($1.05-1.26
\times10^{22}$ cm$^{-2}$).

The temperature distributions for the cool and hot components are
not very similar. It seems that the higher temperature of the hot
components appears at the out rim and Fe-rich regions (c.f. Fig 6).
This is not surprising, since the hot component is responsible for
most of the Fe K emission and also dominates continuum above 4 keV
(Willingale et al. 2002).

The ionization age maps for both components are much similar to the
corresponding ones given by Willingale et al. (2002). We can see
that the map of the hot component is relatively uniform over the
whole remnant, while the cool one shows some structures. The
brightest shell, which has the largest density, doesn't have the
highest ionization age. This leads to the conclusion that the
brightest shell is later shocked. Considering that the cool
component is associated with the reverse shock (Vink et al. 1996,
Willingale et al. 2002), this suggests that the reverse shocked
ejecta may be stratified and thus shocked at different time.

(II) We obtained the spatial distribution of all the elements
concerned. The elemental abundance of each pixel is the mean of the
two components weighted by their emission measures.

Figure 6 displays the abundance maps of Si, Ca and Fe. The contours
have the similar meaning with those in Fig. 4. We can find that Si
and possibly Ca abundances show a jet-like structure in the
northeast and a counterpart in the southwest. This ``jet'' structure
was first suggested by using the Si emission line equivalent width
image (Hwang et al. 2000) and was confirmed with the ratio image of
the Si He $\alpha$ (plus continuum, $1.78\sim2.0$ keV) and
$1.3\sim1.6$ keV (mostly weak Mg He $\alpha$, Fe L plus continuum)
(Hwang et al. 2004). Our abundance maps further confirm their
results. Meanwhile, from the Fe abundance map, we can find that Fe
is relatively poor in the jet area but seems to be rich in the
direction perpendicular to it. It is also clearly shown in the map
that the Fe-rich knots lie outside the Si- and  S-rich ejecta, which
is consistent with the results of Hughes et al. (2000) and
Willingale et al. (2002).

(III) Figure 7 $-$ 9 are the correlation plots of the abundances of
Ne, Si, and Fe with the other elements. S abundance is strongly
correlated with that of Si. Its correlation with other elements are
very similar to those of Si, and are therefore not displayed. We
note that the minimum points are excluded, for they represent
regions which are dominated by the the continuum emission and thus
have no (or very weak) corresponding lines. In order to show the
overall trends of the correlations more clearly, we binned the data
points into 10 channels and overplotted them in the figures with big
pentacles. The corresponding coherent coefficients are given in
Table 1. Apparently Si is well correlated with S, as well as with
Ca. There is also a good correlation between Ne and Mg.

We noticed that in the abundance correlation maps of Si versus Fe,
there seems to be a two phase correlation with the break at Si
abundance equaling to 3. When the Si abundance is lower than 3,
there is a positive correlation with Fe, while negative when higher.
All these features would give us information about the (explosive)
nucleosynthesis of the SN, as will be discussed in detail in $\S$
4.3.

(IV) The Doppler map (c.f. Fig. 4) we got is different from those of
Willingale et al. (2002). They showed that the ejecta in the
Southeast is generally blue shifted, while in the North red shifted.
However, in our map, most regions are red shifted and only a small
fraction in the Southeast are blue shifted. The Doppler map shows
prominent ``bar''-like features from the southeast to the northeast.

We found that the Doppler map is mostly the result of
mis-calibration. The direction of the ``bar''-like feature is
exactly along the CCD columns. Since the CCD pixels in one column
share one read-out circuit, the red shift ``bar'' is probably due to
the inefficiency of the read-out circuit. The generally accepted
velocity for the knots is in the order of 2000 km/s (Willingale et
al. 2002), corresponding to an energy shift of $\sim$15 eV @ 2 keV
(near the Si and S lines). However, the resolution of {\sl
Chandra}-ACIS is about an order of magnitude higher than this value.
The calibration and/or the performance of {\sl Chandra}-ACIS is
probably not good enough to determine such a small energy shift.
Therefore, we conclude that it is very difficult to derive a
reliable Doppler map with the {\sl Chandra}-ACIS data.

\section{Discussion}

\subsection{Jet structure}

The ejecta structure and the explosion asymmetry of Cas A have been
widely studied. It has been long suggested that the core-collapse SN
explosion process is intrinsically asymmetric, from both the
observational and theoretical point of views (Khokhlov et al. 1999
and reference therein). For Cas A, it shows rapidly moving
oxygen-rich material outside the nominal boundary (Fesen \&
Gunderson 1996) and evidence for two oppositely directed jets (Reed
et al. 1999). The aspherical distribution of its ejecta suggests
that the SN explosion, which produced Cas A, may be asymmetric.

Hwang et al. (2000, 2004) have further supported the jet emission of
Si-rich ejecta using the {\sl Chandra} data. These works are based
on the equivalent width (EW) images of the corresponding emission
lines. However the correlation between EW and the abundance can be
distorted by temperature, column density and ionization age. In
other words, the same abundance can be detected as significantly
different EW because of the different line emissivity that is a
function of the local plasma conditions. Willingale et al. (2002)
gave the abundance maps from the {\sl XMM-Newton} data, but the jet
structure is not seen in their results, due to the spatial
resolution limits and their bigger pixel size.

In this paper, we gave directly the element abundances map, among
which Si, S and probably Ca show the jet structure as expected. The
ejecta in the jet area are very enriched in Si, S and Ca, and
relatively poor in Fe (still exists, c.f. Figure 3(B)). This
confirms not only the presence of jet itself but also that these
elements come from incomplete explosive Si-burning rather than
O-burning (KhoKhlov et al. 1999; Hughes et al. 2000). Meanwhile, the
Fe in the counter-jet region (Figure 3(B)) is less enriched than the
off-jet area (Figure 3 (C)), which also implies that the jet
material did not emerge from as deep in the progenitor as other
observed ejecta(Hughes et al. 2000 and reference therein).

\subsection{Iron distribution}

As far as we know, $^{56}$Fe mainly comes from the decay of
$^{56}$Ni, which is synthesized in the (incomplete) explosive
Si-burning. Hwang \& Laming (2003) identified a region of nearly
pure Fe ejecta, which might be a candidate for the site of
$\alpha$-rich freezeout (see also Vink 2004). Early {\sl Chandra}
observation showed that the Fe emission in Cas A is associated with
ejecta and mainly distributed in the faint region outside the bright
Si-dominated shell in the east. The lack of Si beyond the Fe
emission, as well as the lack of Fe inside the Si emission, argues
against projection effects (Hughes et al. 2000).

From our abundance maps, we further confirmed that the ejecta which
is highly enriched of Fe lies outside the Si and S dominated ejecta
in the east. According to the onion model, the lighter an element
is, the larger radius it should be located (Aschenbach 2002). As
long as the projection effect may not play, at least an important
role, this means that the ejecta in Cas A may have suffered some
kind of overturn, either during or after the explosion.  This
overturn might be the result of the neutrino-driven convection
during the initiation of the SN explosion (Hughes et al. 2000).

In the previous {\sl Chandra} results (Hwang \& Laming 2003; Laming
\& Hwang 2003), it has been suggested that the Fe-rich ejecta have
higher characteristic ionization age than the O/Si knots by factors
of a few to 10, implying either a correspondingly higher density or
an earlier shock time. As large density enhancement (or deficits) is
subject to a number of hydrodynamic instabilities that might destroy
the knots within a few shock crossing time scales (Klein et al.
1994; Klein et al. 2003; Wang \& Chevalier 2001; Poludnenko et al.
2002; Poludnenko et al. 2004), such knots are thus not expected to
be seen now. So Hwang \& Laming (2003) claimed that early shock time
and very modest (or no) density enhancement are the most consistent
explanation for the Fe knots that have survived to be seen today.
Comparing Figure 4 (middle panel) and the Fe abundance map in Figure
6, we can find that the Fe-rich knots in the southeast do have
higher ionization age. This further confirms that they might be
early shocked, and might be additinal evidence for the overturn
described previously.

\subsection{Nucleosynthesis}

In the range of about two orders of magnitude, the abundances of Si,
S and Ca show tight correlations with each other (c.f. Fig. 8).
Willingale et al. (2002) also presented the similar results using
the {\sl XMM-Newton} data and suggested this to be strong evidence
for the nucleosynthesis of these elements by explosive O-burning and
incomplete explosive Si-burning due to the shock heating of these
layers in the core-collapse supernova. This kind of good correlation
can also be found between abundances of Ne and Mg, which means that
they might be the ashes of explosive C/Ne burning. All these results
are generally consistent with the (explosive) nucleosynthesis theory
of a massive star (Woosley et al. 2002; Woosley \& Janka 2006).

Fe abundance is not correlated with any other element, with one
exception that there seems to be a possible two phase correlation
with Si (c.f. Fig. 8 or 9). It is contaminated by extremum data
points, for it is difficult to reliably derive the Fe abundances
from Fe-L lines (e.g. Suh et al. 2005). So we picked regions whose
spectra show clearly Fe K lines, and re-draw the Si versus Fe
abundance plot in Figure 10. Now the two phase correlation is more
clearly presented. From Figure 6 we can easily find that the regions
with Si abundance higher than 3 generally concentrate at the
Si-dominated bright shell, which is natural, and also at the jet and
its counterpart. These regions are believed to be the ejecta of
incomplete explosive Si-burning or explosive O-burning. We know that
explosive O-burning would lead to products with little or no Fe
while in the incomplete explosive Si-burning Fe is one of the main
products. From the Si versus Fe correlation plot, we can see that Fe
is not as enriched as Si, but still abundant. This would support
that these ejecta mainly come from incomplete explosive Si-burning,
but also mix with O-burning products. The regions with lower Si
abundance are mainly the relatively faint parts of the remnant and
might be dominated by the shocked circumstellar medium (CSM).  We
note here that the southeast ejecta, which are highly enriched in Fe
and are believed to be synthesized in the complete explosive
Si-burning, are only a small fraction of the total remnant. They
don't affect much to the correlation we discussed above.

\section{Summary}

We did a spatially resolved X-ray spectroscopy of SNR Cas A. An
obvious jet structure can be found in the Si, S and Ca abundance
maps, further confirming the former suggestions. However, the Fe map
shows that it doesn't follow the jet, but distributes somehow
perpendicular to it. Meanwhile, it lies outside the lighter elements
(such as Si and S), which is consistent with the previous results
and might be due to the neutrino-driven convection during the
initiation of the SN explosion. The tight positive correlations of
Si, S and Ca abundances (c.f. Figure 8) suggest that these elements
come from explosive O-burning and incomplete Si-burning. Ne and Mg
abundances also show a good positive correlation, which means that
they should be the ashes of explosive C/Ne burning. The Fe abundance
is positively correlated with that of Si when Si abundance is lower
than 3 solar value, but when the Si abundance is higher, it appears
a negative correlation. We propose this two phase correlation the
result of the ways in which Fe is synthesized. The highly
Si-enriched ejecta concentrated in the jet and the bright shell are
probably a mixture of the explosive O-burning and the incomplete Si-
burning products. The rest part, in contrast, might be dominated by
the shocked CSM.

\begin{acknowledgements}
We acknowledge the use of data obtained by Chandra. The Chandra
Observatory Center is operated by the Smithsonian Astrophysical
Observatory for and on the behalf of NASA. This work is supported by
the National Science Foundation of China through grants 10533020 and 10573017.
\end{acknowledgements}

\begin{table*}
\caption[]{coherent coefficient for data points excluding extremum ones.}
\label{obslog}
\begin{tabular}{cccccccc}
\noalign{\smallskip} \hline \hline \noalign{\smallskip}
Element&O&Ne&Mg&Si&S&Ca&Fe\\
\noalign{\smallskip} \hline \noalign{\smallskip}

O   &  -      &   0.33   &   0.41   &  0.33  &   0.26    &   0.14  &    0.13    \\
Ne  &  0.33   &    -     &   0.49   &  0.19  &   0.16    &   0.05  &    0.25    \\
Mg  &  0.41   &   0.49   &    -     &  0.48  &   0.39    &   0.16  &    0.24    \\
Si  &  0.33   &   0.19   &   0.48   &   -    &   0.86    &   0.35  &    0.23    \\
S   &  0.26   &   0.16   &   0.39   &  0.86  &      -    &   0.33  &    0.11    \\
Ca  &  0.14   &   0.05   &   0.16   &  0.35  &   0.33    &     -   &    0.05    \\
Fe  &  0.13   &   0.25   &   0.24   &  0.23  &   0.11    &   0.05  &     -      \\

\noalign{\smallskip} \hline \noalign{\smallskip}
\end{tabular}
\end{table*}

\begin{figure*}
\includegraphics[width=8cm,clip]{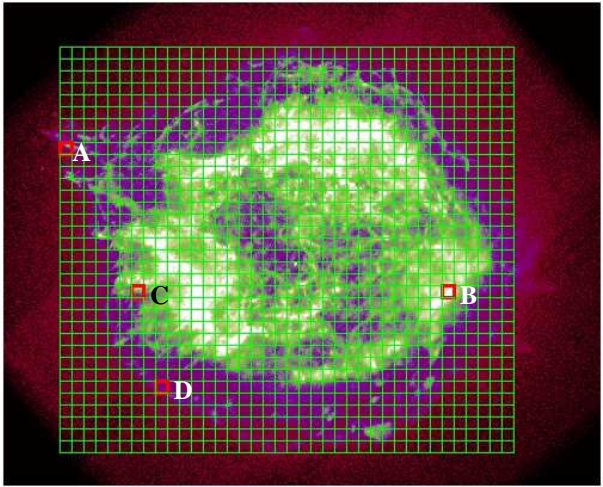}
\caption{The pixel grid used in our analyses superimposed on the
{\sl Chandra} image of Cas A (Observation ID: 4639). Region A, B, C
and D are described in $\S$ 3 and their spectra are presented in
Fig. 3. }
\end{figure*}

\begin{figure*}
\includegraphics[width=6cm,clip]{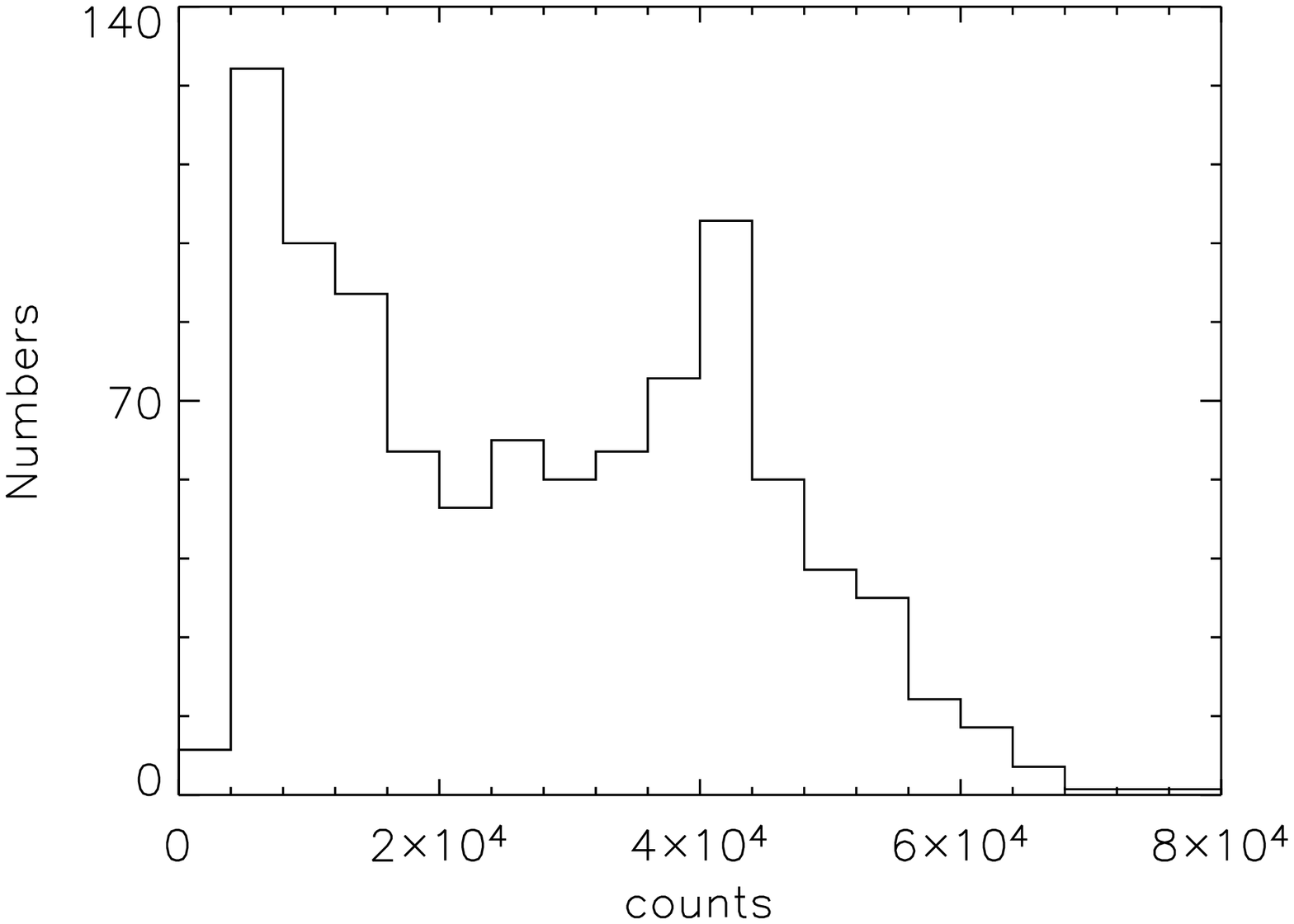}
\includegraphics[width=6cm,clip]{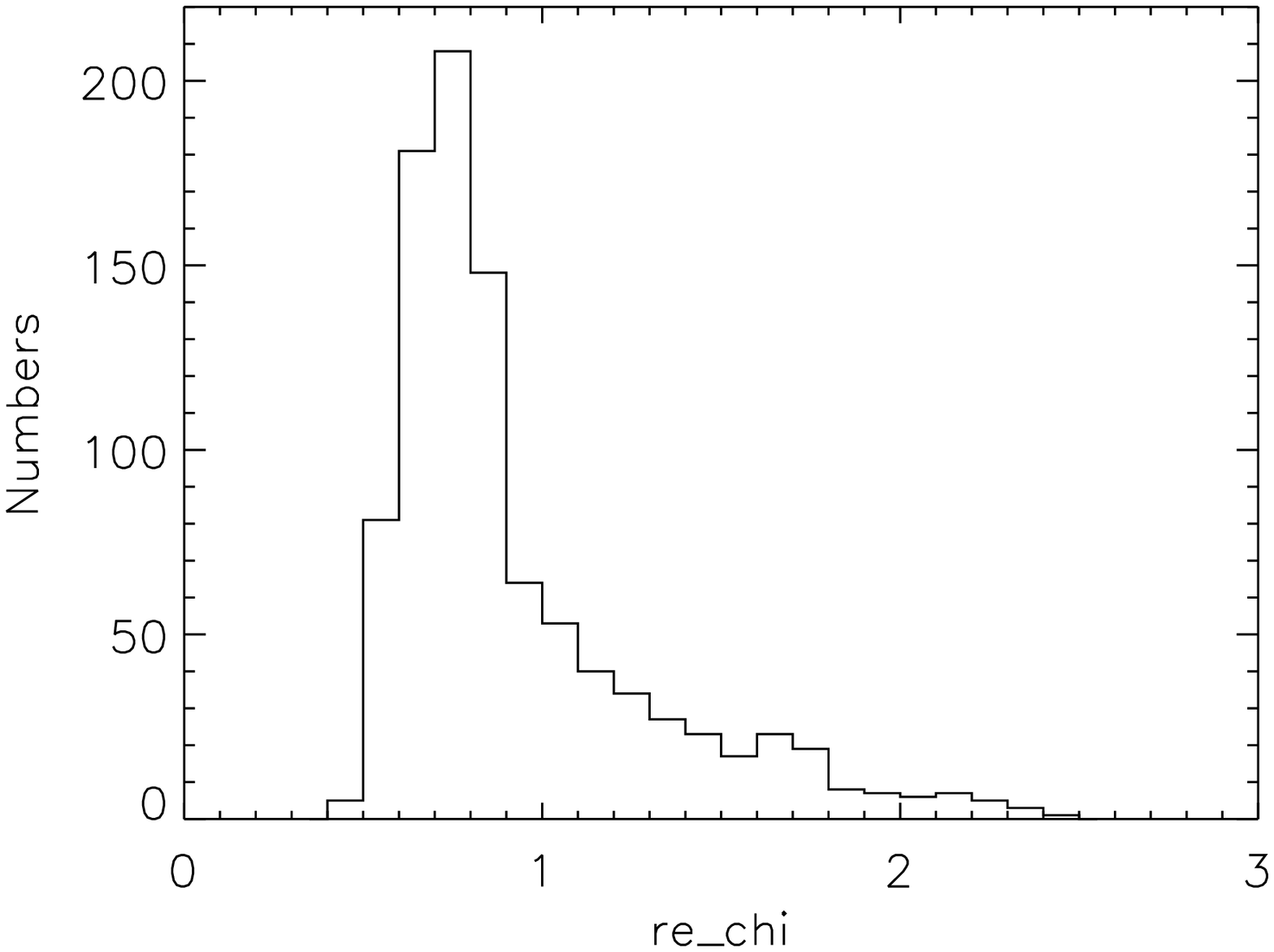}
\caption{Frequency distributions of the 960 pixel counts (left
panel) and ${\chi}^2$ values obtained in the spectra fits (right
panel).}
\end{figure*}

\begin{figure*}
\includegraphics[width=16cm,clip]{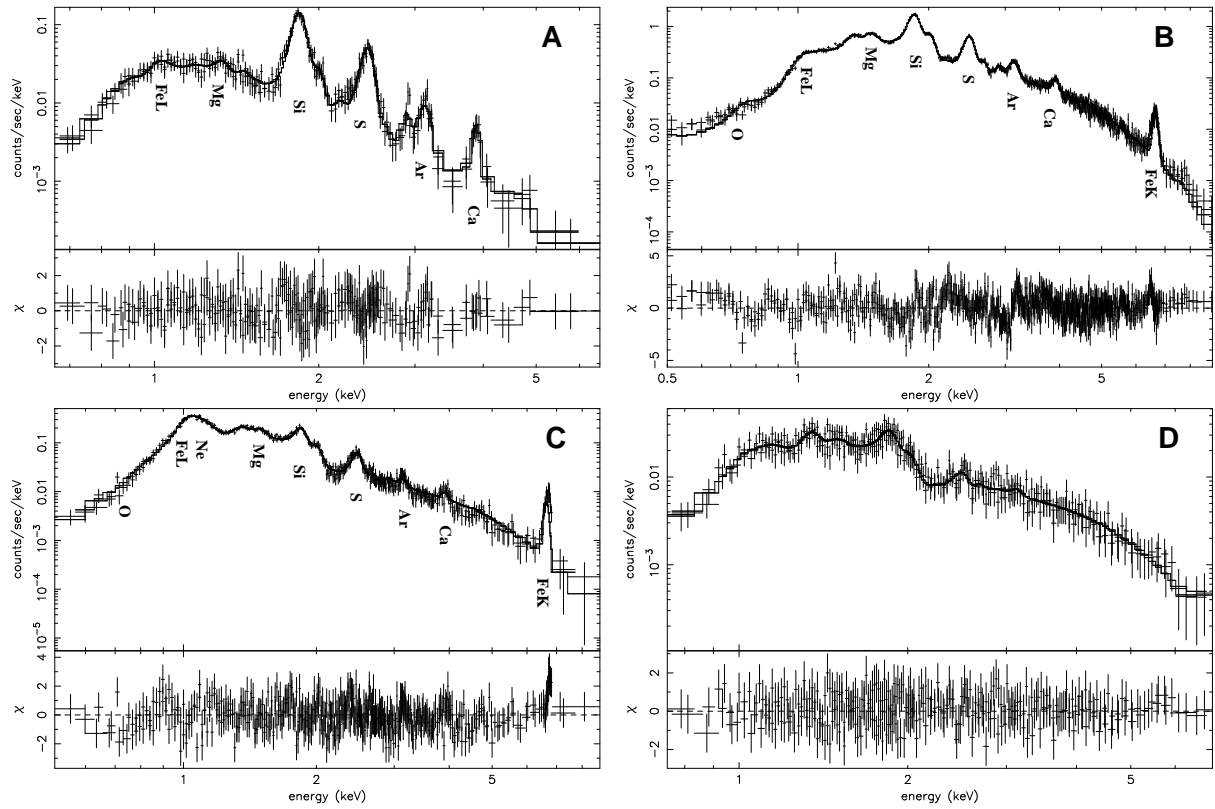}
\caption{Spectra created from regions marked in Fig. 1. Spectrum A represents the spectrum of
the explosive O-burning products, while B and C of incomplete and complete explosive Si-burning
products respectively. Spectrum D is that of the forward shocked ISM. }
\end{figure*}

\begin{figure*}
\includegraphics[width=15cm,clip]{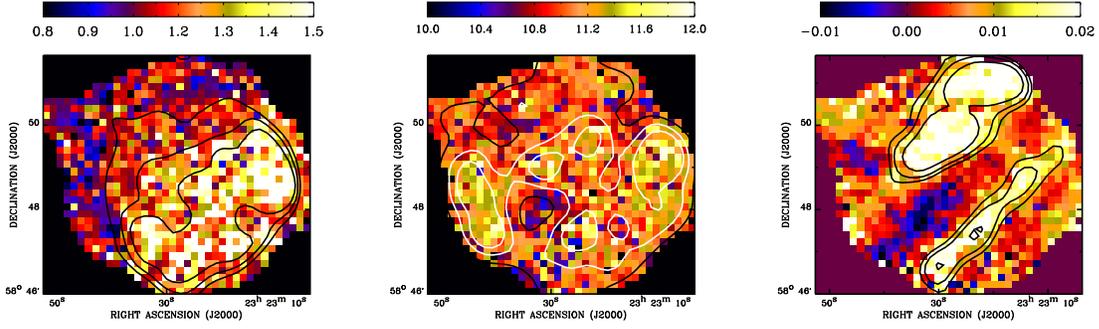}
\caption{The absorbing column density ($N_H$, $10^{22}$ cm$^{-2}$,
left panel), ionization age (Log10 $n_{e}t$ cm$^{-3}$ s, middle
panel), and red shift (right panel) maps of Cas A.
 The negative value of redshift means blueshift. The coding
used is shown on the top of each panel. The overplotted contours
represent the same dataset but smoothed by a two-dimension Gaussian
component with FWHM of $40\arcsec \times 40\arcsec$. The contour
levels are as follows: 1.1, 1.2, 1.3 for $N_H$; 11.0, 11.2, 11.3 for
Log10 $n_{e}t$; and 0.01, 0.012, 0.015 for redshift. }
\end{figure*}

\begin{figure*}
\includegraphics[width=12cm,clip]{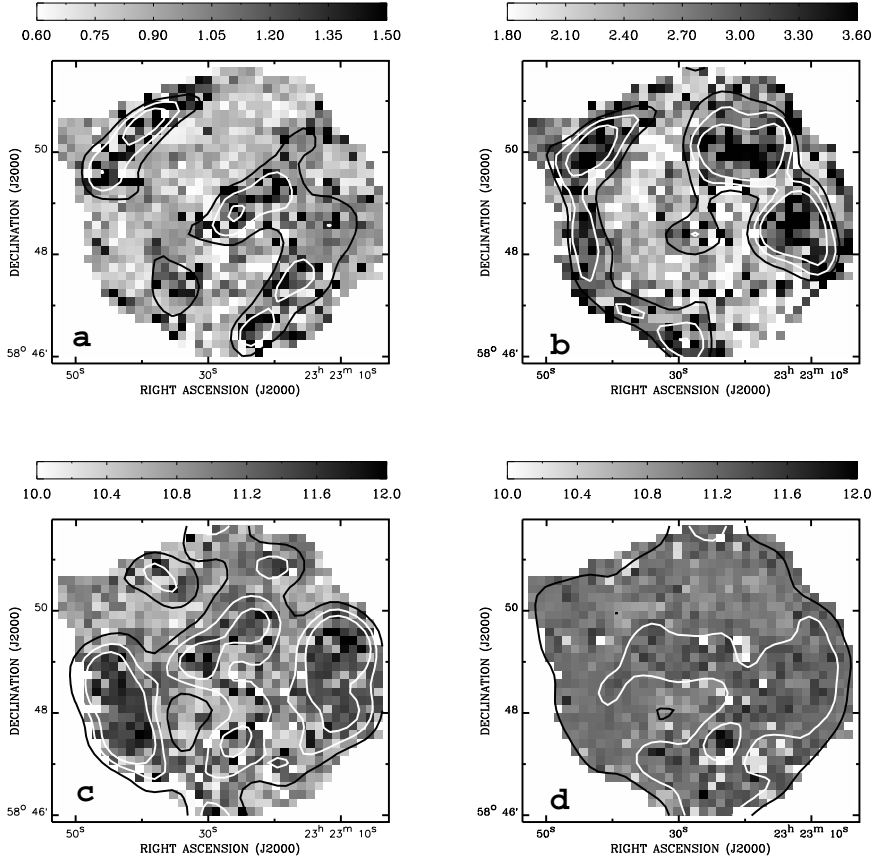}
\caption{The maps of temperatures and ionization ages for the cool
(index 1) and hot (index 2) components of Cas A. a: $kT_1$ (keV), b:
$kT_2$ (keV), c: Log10 $(n_{e}t)_1$ (cm$^{-3}$ s), d: Log10
$(n_{e}t)_2$ (cm$^{-3}$ s). The contours represents the
corresponding images smoothed by the same Gaussian component as used
in Fig. 4. Contour levels are 1.0, 1.1, 1.2 for $kT_1$, and 2.6,
2.8, 3.0 for $kT_2$, while 11.0, 11.2, 11.3 for the two Log10
$(n_{e}t)$. }
\end{figure*}

\begin{figure*}
\includegraphics[width=15cm,clip]{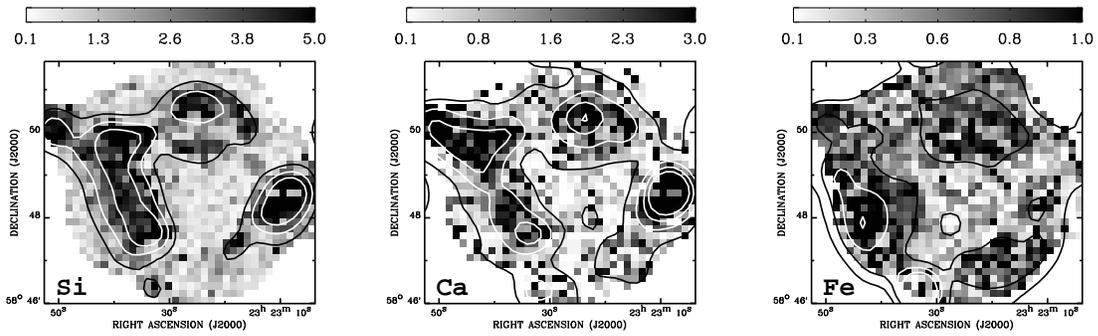}
\caption{Si, Ca and Fe abundance (in units of solar abundance) maps
of Cas A. Contours overplotted represent the same dataset but
smoothed the same Gaussian component as used in Fig. 4. The contour
levels are 2.0, 3.0, 4.0; 1.0, 2.0, 3.0, 4.0; and 0.3, 0.6, 0.9, 1.2
respectively. }
\end{figure*}

\begin{figure*}
\includegraphics[width=6cm,clip]{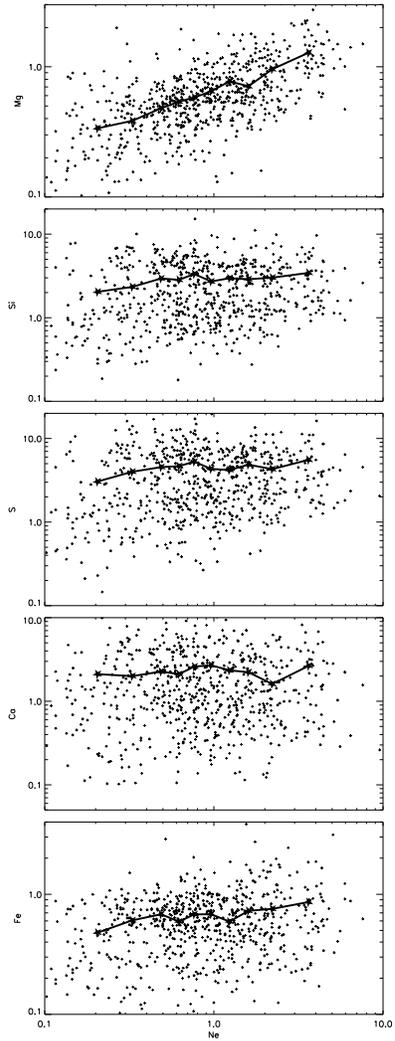}
\caption{The abundance correlation plots between Ne and those of
other elements. All the points were binned into 10 channels and
overplotted in the figure with the big pentacles connected with the
solid line. }
\end{figure*}

\begin{figure*}
\includegraphics[width=6cm,clip]{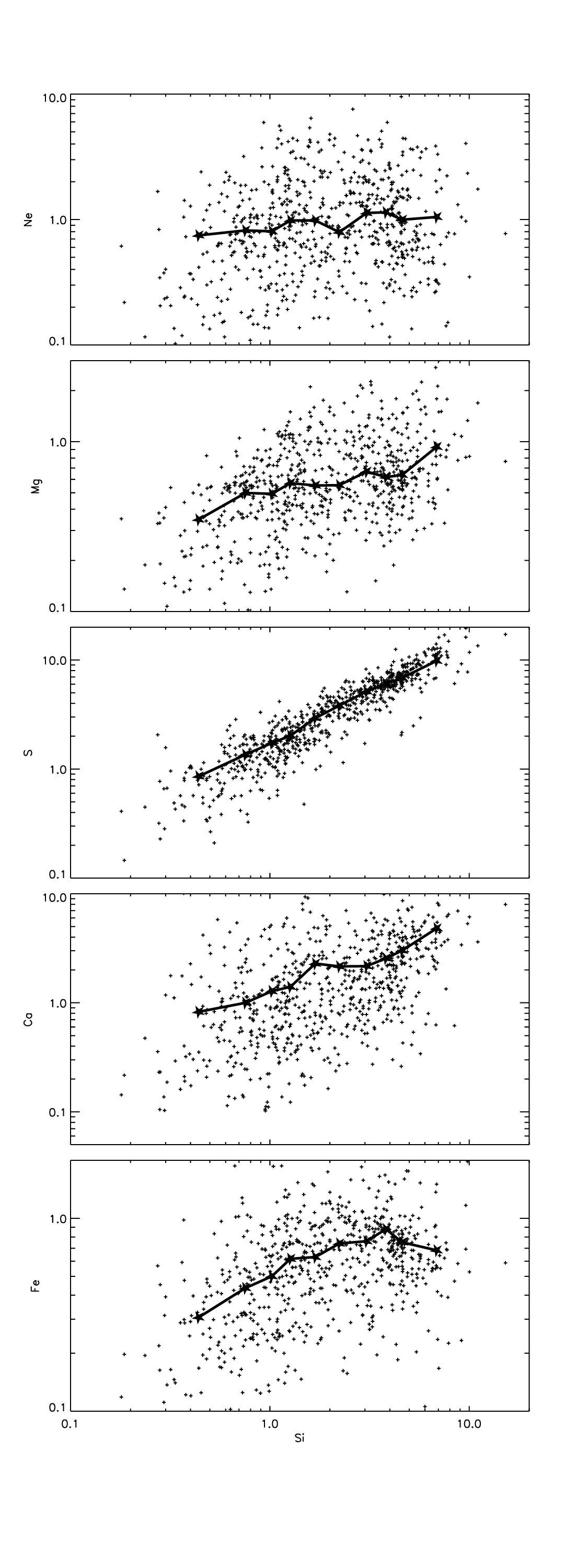}
\caption{The Si abundance correlation plots versus those of other elements.
The big pentacles and solid line have the same meaning as in Figure 7.}
\end{figure*}

\begin{figure*}
\includegraphics[width=6cm,clip]{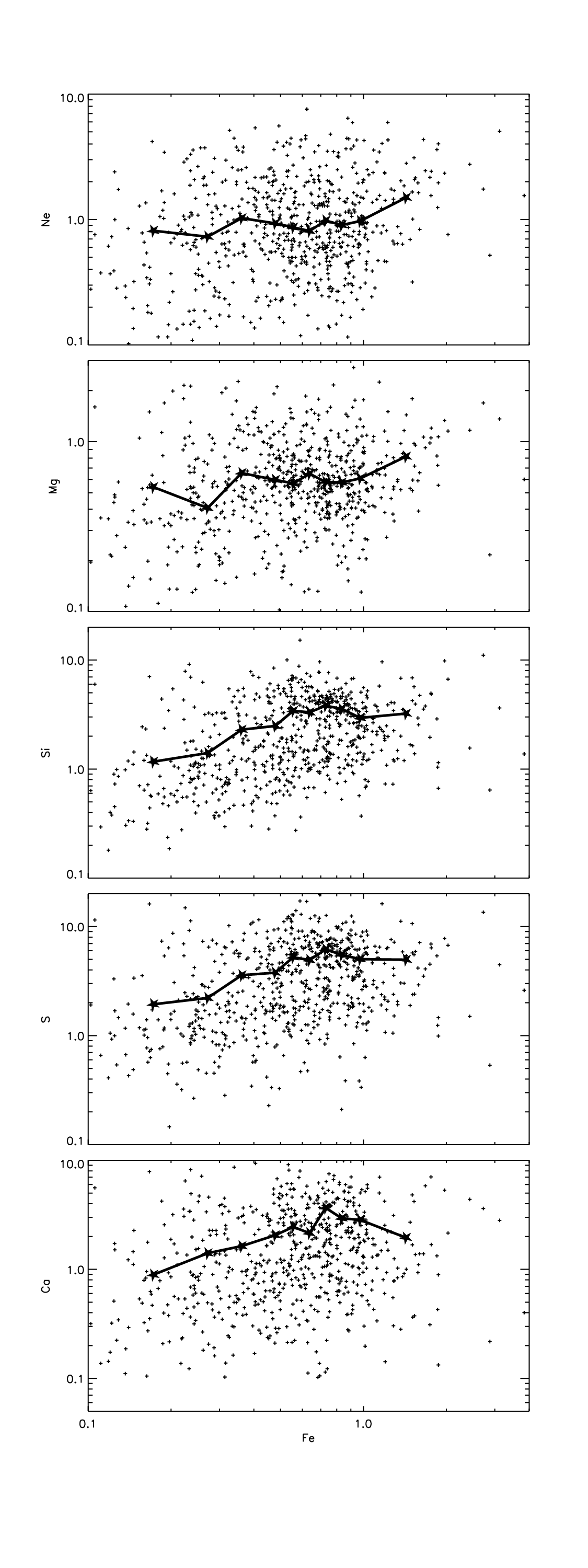}
\caption{The Fe abundance correlation plots versus those of other elements.
The big pentacles and solid line have the same meaning as in Figure 7. }
\end{figure*}

\begin{figure*}
\includegraphics[width=8cm,clip]{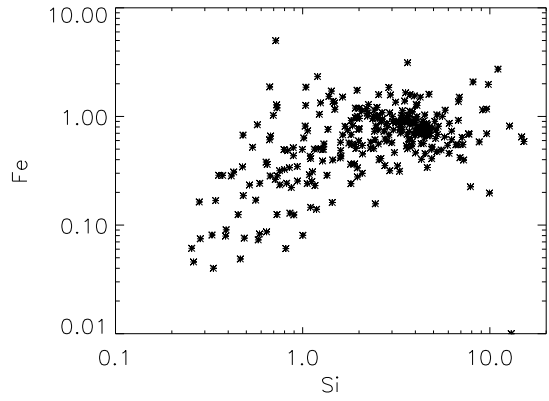}
\caption{The Si versus Fe abundance correlation from regions with
significant Fe-K lines. }
\end{figure*}

\end{document}